\documentclass[11pt,a4wide]{article}
\usepackage{palatino,amsmath,amssymb,graphicx}

\makeatletter

\ifcase \@ptsize
\or 
\or 
\fi
\advance\textheight by \topskip

\ifcase \@ptsize
\or 
\or 
\fi

\begin{document}

\baselineskip 16pt
\parindent 10pt
\parskip 10pt

\noindent{\Large\bf Who has the best probabilities?}\\
\noindent{\bf\large  Luck versus skill in prediction tournaments}

\noindent{\large Niall MacKay,\\ {\em Department of Mathematics, University of York, York, YO10 5DD, UK}}

\vskip 0.1in
\noindent{\bf ABSTRACT}\\
\noindent\parbox[t]{5.5in}{An informal and elementary introduction to probability scoring and forecast verification and improvement, slightly extended from \emph{Significance}  22:3(2025)16, which might be useful  for less mathematical readers as a prologue to the classic review by Gneiting and Raftery [Strictly proper scoring rules, prediction, and estimation, \emph{Journal of the American Statistical Association} 102 (2007): 359]. }

\vskip 0.2in
\noindent{\bf\large
 1. A World Cup Prediction Tournament}

For many years now the mathematics department at York has run prediction tournaments
for the soccer Euros and World Cup finals. Everyone submits their probabilities for each
match -- win, draw, loss -- and is scored on the outcome. I won the first tournament, in 2010, 
just using back-of-the
envelope probability estimates, but did much worse once others started using FIFA and
UEFA data to inform their estimates. 

So, starting from scratch, how do these tournaments
work? How can we create our probabilities, and then judge them against outcomes? Above
all, how can we tell whether the winner truly has the best probabilities, or merely got
lucky? To begin, let’s consider just one (imaginary) match -- let's say Brazil versus Ghana. Six people A-F submit their predictions
for Brazil to win, draw or lose, as in Table 1. (E would like to submit equal
probabilities of one-third for each possibility, but here they are rounded to 0.33.) How
should we score the outcome?

\vspace*{0.2in}
\begin{table}[h]
\centering
\begin{tabular}{|c|ccc|cc|} \hline &&&&&\\
& $p(win)$ & $p(draw)$ & $p(loss)$ & $S_B$ & $S_L$ \\[0.1in]
\hline &&&&&\\
A & 100\% & 0 & 0 & 0 & 0 \\[0.05in]
B & 50\% & 50\% & 0 & -0.25 & -0.69 \\[0.05in]
C & 50\% & 30\% & 20\% & -0.19 &-0.69 \\ [0.05in]
D & 55\% & 45\% & 0 & -0.20 & -0.60 \\[0.05in]
E & 0.33 & 0.33 & 0.33 & -0.33 & -1.10 \\[0.05in]
F & 0 & 100\% & 0 & -1.00 & $-\infty$\\[0.05in]
 \hline
\end{tabular}
\end{table}
\vspace*{0.1in}
\centerline{\parbox[t]{3.2in}{Table 1: Six sets of probability forecasts for one football match, and their Brier and log scores.}}
\vspace*{0.1in}
\vfill\pagebreak
The match is played, and Brazil wins. Whose forecast was best? A's, clearly. But whose is next? We need some function of each person's three submitted forecasts which we can use as a score. One standard solution, a natural one in statistics, is to use the sum of the squares of the differences between the actual outcome (here $(1,0,0$)) and the forecast. We use the negative of this (so that high scores are good), and by convention we halve it.\footnote{We choose our signs so that all scores we consider are `positively oriented': higher scores are better. It is equally possible to use `negatively oriented' scores.}  This is $S_B$, the `Brier score' [1]. For C, for example, $S_B$ is half of $(1-0.5)^2 + 0.25^2 +0.25^2$, or $-0.1875$. Then A, whose forecast was perfect, achieves the highest possible score, of 0. F, whose forecast was perfectly wrong, achieves the lowest possible score, of $-1$. In some sense the base score, the `forecast to beat', is not F's but that of E, who simply gave an equal probability to each possible outcome. Sometimes E is called the `dart-throwing chimp' -- although one could argue that this metaphor is more akin to choosing with equal probabilities (of one third) each of the three forecasts $(1,0,0),(0,1,0)$ and $(0,0,1)$, which would score $-2/3$.

But there is something unsatisfactory about this. C scores better than D, even though C's probability for the actual outcome is only 50\%, lower than D's 55\%. This is clearly because of the way the Brier score handles the two results that did {\em not} occur, and seems rather unfair -- how can we presume to score the probabilities of the outcomes that did not happen?

\vskip 0.2in
\noindent{\bf\large
 2. A Variety of Scores}

There is an alternative, the so-called `log score' $S_L$, which is the natural logarithm\footnote{All logarithms in this paper are natural, to base $e=2.71828...$, but are written `$\log$'. In every case they can, with appropriate changes, be replaced by logarithms to your preferred base.} of the forecast probability for the realized outcome [2].\footnote{Scores which avoid the problem above with C and D's scores are called `local', and that we should not score on unrealised outcomes can be related to the `strong likelihood principle'. The log score is the only possible local score for non-binary outcomes.} Again the best possible score (for A) is 0, but now the worst (for F, with a forecast probability of 0 for the actual outcome) is infinitely negative. This is sometimes viewed as a bad property, but in some circumstances it's accurate: if a gambler stakes their bank on the wrong outcome then they are bankrupt, and out of the game. Both log and Brier score A best, E poorly and F worst. But log now places D second, with B and C equal third. The log score is  natural, because it's the only score which respects conditional probability (giving the same result for multiple independent events whether scored separately or together), and it has some lovely properties. Its expected value (if the forecast probability is correct) is the Shannon entropy. If the gambler bets optimally (`Kelly betting') then the difference between their log score and that of their bookmaker is the realized value of the log of their bank multiplier, and its expected value is known as the Kullback-Leibler divergence (see Box 1).

\vfill\pagebreak
\vspace*{0.5in}
\fbox{
\begin{minipage}{6in}

\centerline{\Large\bf Kelly Betting}

\vskip 0.1in
\centerline{\parbox[t]{6in}{Classic advice to new bettors on horse racing is: don't bet on the favourite, bet on the
horse you think is undervalued; and don't bet your bank, bet in proportion to the
undervaluation. But how much, exactly?

\vskip 0.1in
Imagine the following bet. I toss a coin. I will double your stake on heads, keep it on tails.
(To make some money I might actually merely multiply your stake by a little less than
two on heads, but let's set that aside.) But you know something I don't: you know that
my coin is biased, with a probability $p>0.5$ of heads. We play the game repeatedly. To exploit
this `edge', how much – what fraction $f$ of your `bank' – should you bet on each play?
The answer is clearly not nothing, for then you gain nothing. Nor is it everything, for
then in an average $1-p$ of your plays you will lose everything.

\vskip 0.1in
But this is an easy calculus problem: over $N$ turns, with $pN$ wins and $(1-p)N$ losses, 
my bank is multiplied  by $(1+f)^{Np}(1-f)^{N(1-p)}$ ,
which is maximized at $f=2p-1$ . To bet exactly this much is \emph{Kelly betting}, originally due to
Bernoulli but rediscovered by Kelly at Bell Labs in the 1950s.
The calculus problem is made a little easier by taking logarithms, and (after $N$ drops out)
we have optimized $$p\log(1+f)+(1-p)\log(1-f),$$ which gives an easy way to remember the principle of Kelly betting:
(you should bet so as to) {\bf maximize the expected value of the log of your bank multiplier}.

\vskip 0.1in
In practice, though, Kelly betting is quite risky, for the expected gain rapidly declines if the fractional bet is too large, and thus if
you have overestimated your edge. A common strategy is instead to bet a fixed fraction
of the Kelly fraction, or equivalently to reserve a fixed fraction of your bank as cash.
\vskip 0.1in
Now, when $f=2p-1$ the previous expression becomes 
$$p\log p +(1-p)\log(1-p) +\log 2$$ and is the difference between your and my log score.
This is the Kullback-Leibler divergence
between us, and your log score is (minus) the Shannon entropy.}}

\vspace*{0.1in}
Kelly JL (1956) A new interpretation of information rate, \emph{The Bell System Technical Journal} 35(4): 917--926.
\end{minipage}}

\vspace*{0.2in}

\centerline{\parbox[t]{2in}{\hspace*{0.5in}Box 1: Kelly betting }}

\vfill \pagebreak

\vspace{0.2in}
\begin{figure}[h!]
    \centering
    \includegraphics[width=0.4\linewidth]{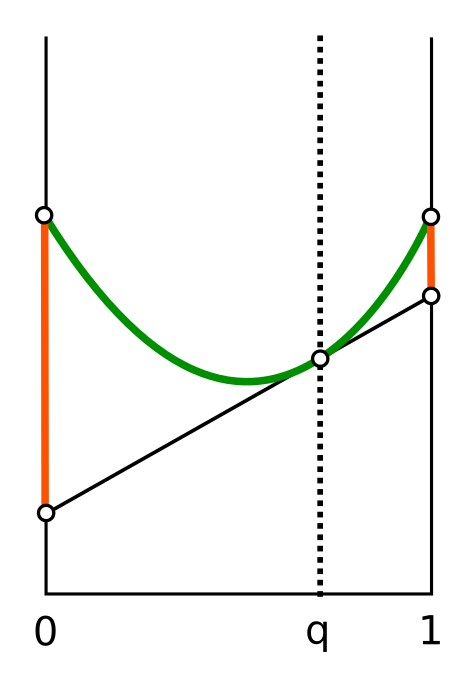}
    \caption{The scoring rule induced by an entropy function $F(p)$. The score is minus the Bregman divergence $D(q,X)=F(X)- \left( F(q)+F'(q)(X-q) \right)$, which is the difference between the function evaluated at the outcome $X=0$ or $X=1$ and a linear approximation to it centred on the forecast probability $q \in [0,1]$. The bold green line shows an example function and the lengths of the orange bars give the sizes of the losses faced by a forecaster who gives probability forecast $q$ when the event in question either does ($X=1$) or does not ($X=0$) happen.}
    \label{fig:savage_rep}
\end{figure}
\vspace{0.2in}

However, the Brier score remains popular, in part because it's easy to work with (as will become apparent below), and the problem described above disappears if there are only two possible outcomes.  In this binary case, how then should we choose a suitable score? In fact infinitely many scores are possible which satisfy the basic requirement of being `proper', which means that their expected value is optimized by giving what you believe to be the true probabilities. A proper score is (at least locally, around the true probability) a concave function of the forecast probability. As a result of this, the score of the average forecast of a group of forecasters  is higher than the average score of their individual forecasts, via Jensen's inequality.\footnote{This states that for a concave (to the origin) function, the function of the average is greater than the average of the function.}

The most general construction (due to Savage [3]) is to begin with any smooth {\em convex} function on the unit interval. This is a (generalized) `entropy', and is the expected value of the score when its argument is the true probability of the event. Convexity is necessary to ensure, again via Jensen's inequality, that a forecaster always benefits from additional information which refines their estimate of the true probability [4]. Proper scoring rules can then be identified with `Bregman divergences' between a forecaster's probability distribution (encoding their beliefs about an outcome) and the degenerate distribution that places all probability mass on the realized outcome. Figure 1 shows this `Savage representation' graphically.

There are three classic simple cases: the Brier score, the log score, and a third, `spherical score' (see Table 2). The best sense of why one might choose a particular score is to consider the entropy and its derivatives. 
The first derivative determines the `exposure', measuring (as in Figure 1) the extent to which the actual outcomes (`true' or `false', 1 or 0) push the score away from its expected value. The second derivative determines the `penalty'. To understand its role, remember that the best possible expected score comes from a forecast which uses the true probability. Suppose the forecaster's probability differs only a little from this.
 In the Taylor series about the true probability, the first derivative, and thus the first order term, vanishes. The forecaster's expected penalty, relative to the score of the true probability, is then approximately given by the second order term,  (half) the square of the difference between the forecast and true probabilities, multiplied by the second derivative. Effectively, this is the sensitivity of the score (the lengths of the orange lines in Figure 1) to small variations in the forecast probability $q$. Box 1 shows how this works for the Brier score, whose second derivative is 2 and thus penalises poor forecasts uniformly across all values of the true probability. In contrast, the log score penalises harshly towards the ends of the interval (recall the bankrupt gambler!) while the spherical score penalizes more in the middle (and so might be good for, say, discriminating slightly-biased coin tosses or other finely-balanced situations). The entropy, exposure and penalty for the Brier, log and spherical scores are given in Table 2.

\vspace*{0.25in}
\begin{table}[h]
\centering
\begin{tabular}{|c|ccc|}\hline &&&\\
Score:& Brier & log & spherical   \\[0.05in]
\hline &&&\\
entropy & $p(p-1)$ & $p\log p +(1-p)\log(1-p)$ & $r$ \\[0.05in]
exposure & $2p-1$ & $\log\left({p\over 1-p}\right)$ & ${2p-1\over r}$ \\[0.1in]
penalty & $2$ & ${1\over p(1-p)}$ & ${1\over r^3}$ \\[0.15in]
 \hline
\end{tabular}
\end{table}
\vspace*{0.1in}
\centerline{\parbox[t]{3.6in}{Table 2: The most commonly used proper scores.  The spherical score is written in terms of $r^2=p^2+(1-p)^2$. }}
\vspace*{0.25in}

\vskip 0.2in
\noindent{\bf\large
 3. Building Better Forecasts}

But how should we improve our scores, in practice? We need to understand how to
construct our forecast probabilities, the different features the construction might
have, and how good we are at each of those features -- that is, we need some metrics for
our forecasting strategy. The Brier score is especially amenable to analysis in this
respect, because it is easily decomposable into different contributory parts. Similar
decompositions are possible for other scores,
but are a little more complicated.

Suppose you've played a tournament of simple binary forecasts (such as `will it rain here tomorrow?'), you have your record available to analyse, and you want to understand how you might improve your forecasting. You do not and can never know the `true' probabilities, of course. Suppose further that your method was to make your forecasts in bins, of events whose probability you think you can't reliably tell apart. For example, you might just have five bins, for events which you think have probabilities of 0--20\% (for which you forecast 10\%), 20--40\% (forecast 30\%), 40--60\% (forecast 50\%), 60--80\% (forecast 70\%) and  80--100\% (forecast 90\%), and put each event to be forecast in one of these bins. Then the average Brier score can  be decomposed into three elements:  an intrinsic and unavoidable `uncertainty', determined by the actual overall frequency of `true' outcomes, and which you can't alter;  a `reliability' measure, of the accuracy of your bins (how close to 30\% was the proportion of `trues' among forecasts in the 20-40\% bin?); and a `resolution', which is improved by having more bins (could you have done better by refining into 10 bins?).\footnote{Reliability and resolution can also be considered together using the `receiver operating characteristic' (ROC) curve.}

The decomposition is
\[
\begin{array}{cccccccc}
S_B & = &   - & f(1-f) &  +&\frac{1}{N}\sum_\mu N_\mu(f-f_\mu)^2 &-& \frac{1}{N}\sum_\mu N_\mu (q_\mu-f_\mu)^2 ,\\[0.1in]
& =& -& \mathrm{uncertainty} & +&\mathrm{resolution} & - & \mathrm{reliability}
\end{array}
\]
where $f$ is the actual frequency of true outcomes among $N$ events, the bins are labelled by $\mu$, $f_\mu$ is the frequency of `trues' among $N_\mu$ events in that bin, and $q_\mu$ is the forecast associated with that bin. The proof first expands the brackets (using $N=\sum_\mu N_\mu$ and $Nf=\sum_\mu N_\mu f_\mu$) to give
$$
S_B =   \frac{1}{N} \sum_\mu N_\mu(- f_\mu+  2q_\mu f_\mu -q_\mu^2 ).
$$
But the average Brier score itself, defined using the sum over all events $i$ of the event $X_i(=0$ or $1)$ with forecast $q_i$ (and noting that $X_i^2=X_i$), is 
$$
S_B=-\frac{1}{N}\sum_i (X_i-q_i)^2 =\frac{1}{N}\sum_i (-X_i+2q_iX_i - q_i^2),
$$
and the two expressions are the same when we note that $\sum_i X_i = \sum_\mu N_\mu f_\mu$.

This is the classic approach [5], but others are possible, according to your forecasting method and context.  For example, consider (as a binary, win/loss) a match between two football teams who meet regularly -- say Arsenal versus Everton. We might try to place it within a bin (`This should be in my approximately-70\% win bin'), and then the decomposition above is the way to analyse results. But we might instead begin by looking at the historical frequency of results of this match, and then try to refine this by looking at the current teams. If we look at how our method works on training data, for which the historical frequency $f$ is known, and our method refines this to $q_i=f+\epsilon_i$, then
\begin{eqnarray*}
S_B & =&  -\frac{1}{N}\sum_i(X_i-q_i)^2 \\
& = & -\frac{1}{N}\sum_i X_i -2X_i(f+\epsilon_i) + (f+\epsilon_i)^2 \\
& = & -f(1-f) +\frac{2}{N}\sum_i \epsilon_i(X_i-f) - \frac{1}{N}\sum_i\epsilon_i^2
\end{eqnarray*}
The first term is the base score, the intrinsic uncertainty, just as before. This is the score to beat! The other two terms together score your departures $\epsilon_i$ from $f$. You are penalised for these by $\sum \epsilon_i^2$, but you will also gain if your $\epsilon_i$ are correlated with results $X_i-f$. You could think of the $\sum \epsilon_i^2$ as what you stake in order to achieve a payoff $\sum \epsilon_i(X_i-f)$. Indeed, one can make this precise: suppose we write $\epsilon_i=R\gamma_i$ with $\sum\gamma_i^2=1$. That is, you allow ourself total $\sum_i\epsilon_i^2=R^2$, while your prediction model's job is to decide on the relative values of the $\gamma_i$. Then the optimal $R=\sum_i\gamma_i(X_i-f)$, and this tells you how far to back your model. If $\sum_i\gamma_i(X_i-f)<0$, your model's predictions are anticorrelated with reality, and you should probably simply forecast the historical frequency $f$ and go down to the pub to watch the match.

Using historical frequencies is something like the approach of the lazy forecaster E in our opening, trinomial example, who could easily refine their uniform thirds into the historic proportions of wins, losses and draws. Indeed there is usually a baseline `(lazy) forecaster to beat', and the Brier score is often refined into a `skill score'. At its simplest this is the difference between the forecaster's score and that of the uniform forecaster E, divided by the score of E -- here, $3B+1$ -- so that only a positive score indicates genuine skill. Such a score is proper, because a linear function of a proper score is also proper.

However, we might want instead to use the actual historical frequencies in the comparison score, 
since different frequencies of events will give rise to different baseline uncertainties. For example, in a place where it typically rains on half of all days, the basic (binary) score to beat will be $-0.5$, whereas in a much drier place where it rains only on one day in ten this baseline score will be $-0.1\times 0.9=-0.09$. To use $B/0.09 +1$ might then be thought to enable us to compare forecasts in the two locations.
However, if we rescale in this way then the penalty, and thus the variation in final scores due to forecaster skill, will also be rescaled. 
Further, the `skill score' would become improper if we were to score both forecaster and a non-uniform comparator on the actual events.  In the end there is no fair way to compare scores in different tournaments, for different sets of events with different observed frequencies.
Beyond this, a nonlinear function of a proper score is generally improper -- so if you want, say, to reward a weather forecaster using their Brier score, then their bonus must be proportional to their score, rather than being triggered by its exceeding some threshold. 

Suppose now that we want to predict whether an exceptional event will occur -- an earthquake, say, or a volcanic eruption -- and reward the forecaster in a manner which incentivizes them to give their best estimate of the true probability.  This means that we certainly want to treat `true' and `false' events very differently, which the Brier, log and spherical scores do not do: they are all invariant under the exchange of `true' and `false', or equivalently under inserting a negation into the forecast question.  It is easy to construct asymmetric scores -- one can simply include a payoff that depends on the event but not on the forecast. But it is also possible to construct
truly asymmetric proper scores in a way that treats the `true' and `false' forecast probabilities differently.\footnote{In an Appendix we provide an asymmetric generalization of the spherical score which we call the `elliptical score', and which is the only point in this article which does not appear to be in the literature.}

For our earthquake forecaster, one possible proper asymmetric score (related to the Poisson distribution) rewards the forecaster by giving them a reward proportional to their probability for `false' whether there is an event or not, and adding to this the logarithm   (which is negative, and is thus a penalty) of their probability for `true' when an event  does occur -- that is, their score for forecast $q$ is $X\log q + 1-q$. (We probably wish to disallow a forecast of $q=0$, in what statisticians sometimes call `Cromwell's rule'.\footnote{This is the idea that everything should be allowed to be possible, to be conceivable, so that one should never offer perfect probabilities of 0 or 1. It has its origins in English Republican leader Oliver Cromwell's plea to the Scottish army before the battle of Dunbar, `I beseech you, in the bowels of Christ, think it possible you may be mistaken'. As with all Cromwell quotations, this is best spoken in a broad Fenland accent, easily accessible only a few miles from Cambridge but rarely heard by its academic denizens.})
For example, suppose that most of the time the correct forecast is below one in a million. Almost all of the time the forecaster gets their normal payment minus this tiny proportion. If the forecaster raises the forecast to one in ten then they forfeit almost a tenth of their usual payment -- but if an earthquake were to occur then they would further forfeit only the log of ten, rather than the log of a million.

\vskip 0.2in
\noindent{\bf\large
4. Luck versus Skill}

Now let's return to our prediction tournament. In the light of what we've learned, we shouldn't have a sweepstake, a winner-takes-all prize. Remember that if the reward is a nonlinear function of a proper score (such as a step function) then the reward is not maximized by submitting  true probabilities. What would happen instead is that the forecasters standing second or third in the rankings approaching the last few games would stop giving their true probabilities and start giving more definite forecasts -- for example, forecaster D might believe their (55\%, 45\%, 0) to be the true probabilities, but actually predict the (100\%, 0, 0) of forecaster A, taking a chance in the hope of winning.  Alternatively
… well, in our first tournament, I was in the lead with a few games to go. Coming second was my then head of department, an eminent pure mathematician who shall remain nameless. So, to make sure that I could not lose, I used much more cautious probabilities for the last few games than those I had calculated – and realised that these were almost exactly the bookmakers' probabilities. It turned out that the head had simply been submitting probabilities inferred from bookies’ odds throughout. So, even if the score used is proper, to incentivise players to submit their true estimates of the probabilities all forecasts need to be submitted in advance

But there is still the fundamental question of luck versus skill. Is the winner the best forecaster, or just a forecaster who got lucky? In the long run the central limit theorem will do its work -- scores will bunch closely, and skill will become more important relative to luck. After all, in a tournament of $N$ predictions we are dealing with binomial distributions, whose large-$N$ limit is classically a bell curve by the De Moivre-Laplace theorem. 

But we really do need to know just how long this takes, and what the balance of skill and luck looks like. A recent paper by David Aldous [6] conducts some simulations to investigate this point, but we can say a little more by using the following identity. Unlike in the previous section, we assume that we do indeed know the true probability, so that ${\mathbb E}X=p$. Then for a single event
\vspace*{0.1in}
$$
\begin{array}{ccccccccc}
S_B(q) & = &-(X-q)^2 & = & -p(1-p)  & + & (2q-1)(X-p) & - &(p-q)^2  \\[0.1in]
&&&& {\mathrm{convex\; in}}\; p &&&& {\mathrm{concave\; in}}\; q\\[0.1in]
&&& = & {\mathbb E} S_B(p) & + & S_B(q)-{\mathbb E}S_B(q) & + & {\mathbb E}(S_B(q)-S_B(p))\\[0.1in]
&&& = & {\mathrm{entropy}} & + & {\mathrm{exposure}} & - & {\mathrm{penalty}}\\[0.1in]
&&&  = &{\mathrm{uncertainty}} & +  &{\mathrm{luck}} &  + & {\mathrm{skill}}
\end{array}
$$

\noindent 
Taking the three terms in turn,

--- the first, convex `entropy' term depends only on the true probability $p$ -- which, of course, is not and never can be known. It is the expected score of the forecaster who knows the true probability -- let's call them the `savant'. 

--- the second, `exposure' term is the only one of the three that depends explicitly on the outcome, the value taken by $X$. Notice that if the forecast is $0.5$ then the exposure is zero, and the forecaster scores $-0.25$ every time, whatever the outcome. The variance of such a forecast is then zero: this is a forecaster for whom there is no such thing as luck, analogous to forecaster E in our opening example.  For other forecasters the variance of the exposure is  $(2q-1)^2p(1-p)$, and in the usual manner for the average score reduces over many questions by a factor of their number.

--- the third,  `penalty' term is the expected deficit in score compared to that of the savant. It depends on both the true and the forecast probability, and is zero when these are equal, enabling us to see immediately that the Brier score is proper -- that it is maximized when the forecast probability is the true probability. However, its being maximal also means that the score is flat here, so that the penalty is second-order and therefore very small.
For example, a forecast error of 10\% results in a penalty of $-0.01$. Suppose the true probability is $0.5$, scoring $-0.25$. We can see that among the Brier scores in our tournament only a tiny proportion of the the variation will be due to skill.  

Now, the balance between luck and skill is precisely that between exposure and penalty.
The exposure naturally has a much greater effect early in the tournament. With the values used in the previous paragraph, its variance is $0.01$, and the standard deviation $\sigma$, the natural measure of typical variation, is $0.1$, ten times larger than the penalty of $0.01$. It thus takes 100 questions before $\sigma$ is reduced to $0.01$ -- that is, before luck outweighing skill becomes a one-$\sigma$ event, occuring with a frequency of about 16\%. It takes 400 questions before it becomes a two-$\sigma$ event, occurring with a frequency of about 2\%. In this tournament, our 10\%-off-forecaster still has a 16\% chance of beating the savant over 100 questions, and a 2\% chance of beating them over 400.

%
%
%
%
However, remember that the savant is fictional: we can never know the true
probability for each question. We need instead to compare two forecasters directly [7].
To do this we need
$$
S_B(q)-S_B(q') = -(X-q)^2+(X-q')^2 = (q'-q)(q+q'-2X)
$$
and its average over $N$ questions, which is
$$
\Delta:=\frac{1}{ N} \sum_{i=1}^N (q'_i-q_i)(q'_i+q_i-2X).
$$
Now, $\Delta$ is approximately normally distributed with mean 
$$
\frac{1}{N} \sum_i (q_i'-q_i)(q_i'+q_i-2p_i)
$$
and variance
$$
\sigma^2={1\over N^2}\sum_{i=1}^N 4p_i(1-p_i)(q'_i-q_i)^2.
$$
Of course we do not know $p_i$, but $4p_i(1-p_i)\leq 1$ (and indeed for $p_i\simeq0.5$ is quite close to 1 over a wide range), so
$$
\sigma^2 \leq {1\over N^2}\sum_{i=1}^N (q'_i-q_i)^2.
$$
So two forecasters who have a root-mean-square difference of $\delta$ in their forecasts will have a standard deviation of less than $\delta/\sqrt{N}$ in the difference $\Delta$ of their average Brier scores.

Thus the difference between two forecasters’ scores is normally distributed, with a variance that
is difficult to estimate (without knowing the true probabilities) but easy to give an upper bound for. 
The hypothesis that there is no difference in the two forecasters’ abilities
begins to look highly implausible -- say, beyond two standard deviations into a tail of
the distribution, and thus with at least 95\% confidence -- when, over 100 questions, and
for forecasters whose probabilities typically differ by about 0.1, their mean scores differ
by more than 0.02. Such a difference in scores may seem small, but remember that typical
scores will be around –0.25.

Finally, let’s apply this (suitably generalized) to the trinomial outcomes and 64 matches of our original FIFA World Cup tournament. The uniform forecast baseline score would be –0.333. If
we refine this using historical frequencies
(to around 0.24 for a draw and 0.38 for either
side to win) we get –0.327. Forecasts were
submitted in rounds, and you can see the
evolution of different forecasters’ individual
average scores in Figure 2.

\centerline{\includegraphics[keepaspectratio=true,
width=7in]{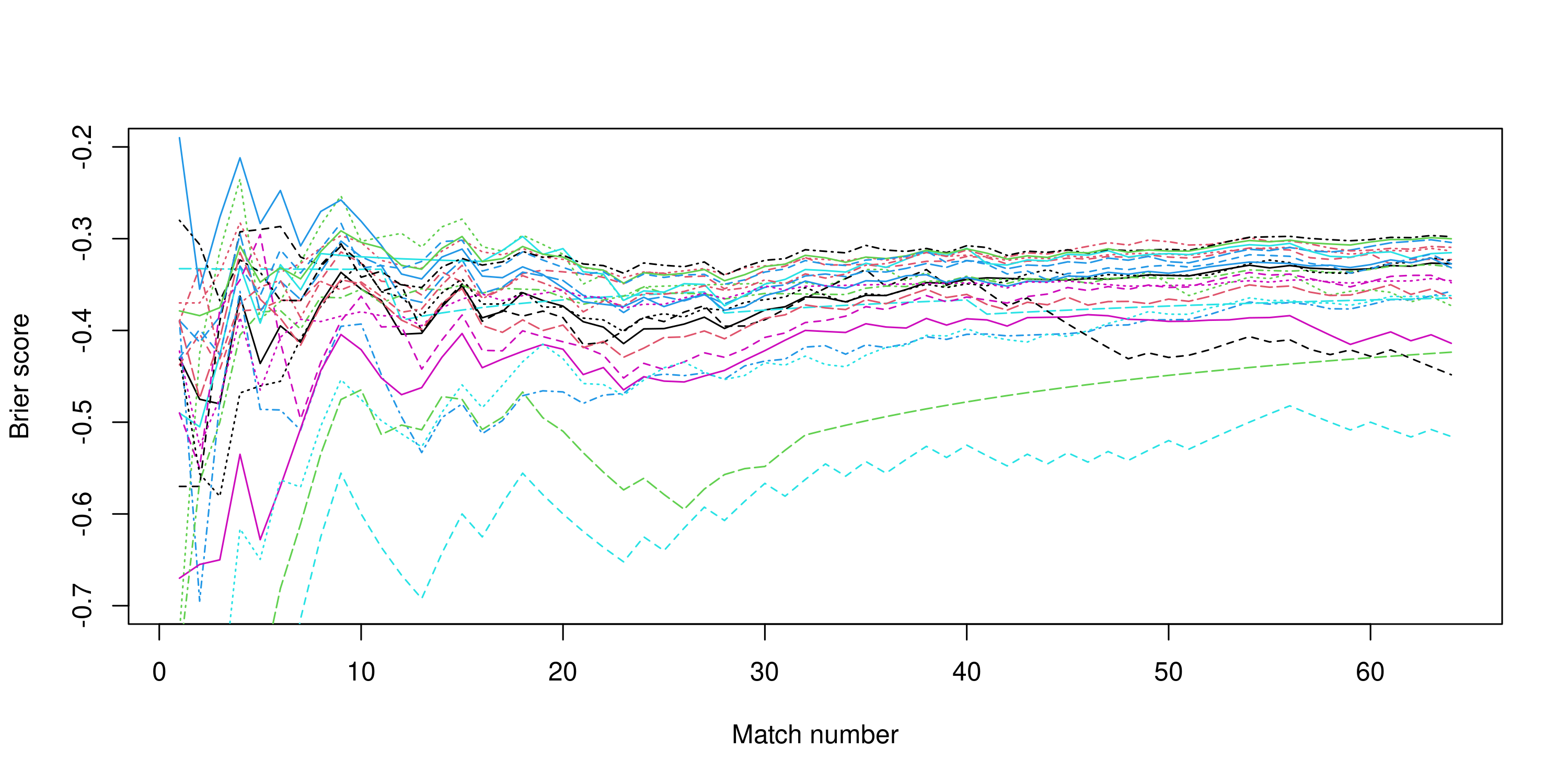}}

\centerline{\parbox[t]{5.5in}{Figure 2: The evolution of forecaster average Brier scores over the matches of the 2010 FIFA World Cup finals. Most scores and differences stabilize, although some players are clearly inclined to gamble -- which is not a strategy for final success!}}

\vskip 0.2in
There is a clear sense that the scores and their differences are stabilising over time
(although there are clearly a few gamblers). Whether the winner really is the best forecaster will of course depend on how
closely players bunch behind the winner, but a rule of thumb might be that the winner is
almost certainly also the best forecaster if they win by a margin of over 0.02, roughly two standard deviations. Well, my winning
score was –0.298, with a margin of 0.002, and in the two tournaments in which we used
the Brier score the gap between first and third place was rather less than 0.01. So in
each case the winner had probably got lucky. More broadly, the top 15–20 forecasters were separated by less than 0.1 (and about half of
these did worse than the uniform baseline). They could be ranked confidently only
within (a two-$\sigma$ variation of) about 3–5 places
either way.

After 2014, for the reasons outlined above, we switched to using log scores. This had
the additional amusing benefit of seeing players go all-in towards the end of the
tournament … and finish with a score of minus infinity, equivalent to losing their
skin. Is it ever possible to make money? Recall that my winning margin of 0.002
was over a second-placed player who used bookmakers’ odds. This was just too small,
and too luck-dependent – and anyway less than the bookies’ margin. So perhaps the
lesson is that you can lose (-- everything, if your
luck is bad) but you are highly unlikely to
make money. 

\parindent 0pt
\vspace*{0.1in}
{\bf Acknowledgments}

The classic review of this topic, from which I learned much and which is in no way responsible for anything I failed to learn or misapprehended, is Gneiting and Raftery [8]. I should like to thank my York colleagues Ben Powell, Stephen Connor and Jamie Wood for discussions, and Stephen Connor for providing Figure 2.

\parskip 8pt
\vspace*{0.2in}
{\bf\large References}

[1] Brier GW (1950) Verification of forecasts expressed in terms of probability, \emph{ Monthly Weather Review}78(1): 1--3.

[2] Good IJ (1950) Rational decisions. \emph{Journal of the Royal Statistical Society} B14(1): 107--114.

[3] Savage LJ (1971) Elicitation of personal probabilities and expectations. \emph{ Journal of the American Statistical Association} 66(336): 783--801.

[4] McCarthy J (1956) Measures of the value of information. \emph{Proceedings of the National Academy of Sciences} 42(9): 654--655.

[5] Murphy AH (1973) A new vector partition of the probability score, {\em  Journal of Applied Meteorology and Climatology} 12: 595--600.

[6] Aldous D (2021) A prediction tournament paradox, {\em  The American Statistician} 75: 243--248. 

[7] Lai TL, Gross S and Shen D (2011) Evaluating probability forecasts, {\em Annals of Statistics} 39: 2356--2382.

[8] Gneiting T and Raftery A (2007) Strictly proper scoring rules, prediction, and estimation, {\em Journal of the American Statistical Association} 102: 359--378.

\vfill\pagebreak
\parindent 10pt
\parskip 10pt

\noindent{\Large\bf APPENDIX: The Elliptical Score}

We noted earlier that it is straightforward to generalize the spherical score (for binary outcomes) to  an `elliptical score'. Since we have not been able to find this in the literature, we include it here.

The elliptical score of a Bernoulli trial $X$ is defined by entropy
$$E_\alpha(p) := \frac{r}{\sqrt{\alpha(1-\alpha)}}, $$
for some $0<\alpha<1$, with exposure
$$E_\alpha'(p)=\frac{p-\alpha}{\sqrt{\alpha(1-\alpha)}}\,\frac{1}{ r}$$
and penalty
$$E_\alpha''(p)=\frac{\sqrt{\alpha(1-\alpha)}}{r^3},$$
where $r^2=(1-\alpha) p^2 + \alpha(1-p)^2$.
For a forecast $q$, events $X$ score
$$
E_\alpha(q)+(X-q)E_\alpha'(q)=X \sqrt{\frac{1-\alpha}{\alpha}} \left(\frac{q}{r}\right) \\[0.1in]
+(1-X) \sqrt{\frac{\alpha}{1-\alpha}} \left(\frac{1-q}{r}\right).
$$

The spherical score is the special case $\alpha=\frac{1}{2}$, and its corresponding scores $\frac{q}{r}$ for $X=1$ and $\frac{1-q}{r}$ for $X=0$ can be thought of as the projection (from the origin) of the line segment $\{(q,1-q)\,\vert\, 0\leq q \leq 1\}$ onto the unit circle. The elliptical score is then the projection of the line segment $$
\left\{\left({\sqrt{1-\alpha\over \alpha}}q,{\sqrt{\alpha\over1- \alpha}}(1-q)\right)\,\vert\, 0\leq q \leq 1\right\}
$$ onto the ellipse $(\alpha x^2 + (1-\alpha) y^2=1$.

Just as the spherical score is most sensitive to inaccurate forecasts around $q=\frac{1}{2}$, so the elliptical score is most sensitive to inaccurate forecasts around $\alpha$. One could use it, for example, as a dynamical parameter in prediction tournaments, tracking crowd wisdom $\hat{p}$, the average forecast of a group of forecasters. Then $E_{\hat{p}}(q)$ is indifferent between outcomes at $q=\hat{p}$ (since $X=0$ and $X=1$ alike score $1$), favours correct opposition to crowd wisdom (in the sense that the $q$-forecaster's departure from crowd wisdom towards the correct outcome $\hat{p}$ scores more highly when the crowd favoured the opposite outcome), and discriminates best when closest to the centre of the crowd.

\end{document}